
\input harvmac
\noblackbox
\def\Title#1#2{\rightline{#1}\ifx\answ\bigans\nopagenumbers\pageno0\vskip1in
\else\pageno1\vskip.8in\fi \centerline{\titlefont #2}\vskip .5in}

%
%
\def\p{\partial}

\def\vp{\varphi}
\def\({\left (}
\def\){\right )}
\def\[{\left [}
\def\]{\right ]}

\def\ajou#1&#2(#3){\ \sl#1\bf#2\rm(19#3)}

\def\phiT{{\widetilde \varphi}}
\def\hB{{\widehat B}}
\def\hq{{\widehat q}}
\def\frac#1#2{{#1 \over #2}}

\def\p{\partial}

\def\R{\hbox{\rm I \kern-5pt R}}

\def\l3{\lambda_3}

\hyphenation{par-am-et-rised}

%
%
\lref\QEP{T.C. Bradbury,\ajou Ann. Phys. &19 (62) 323;
D.G. Boulware,\ajou Ann. Phys. & 124 (80) 169;
P. Candelas and D.W. Sciama,\ajou Phys. Rev. &D27 (83) 1715.}
\lref\brill{D. Brill and H. Pfister, \ajou Phys. Lett.  &B228 (89) 359;
D. Brill and G. Horowitz,\ajou Phys. Lett. &B262 (91) 437.}
\lref\BDDO{T. Banks, A. Dabholkar, M.R. Douglas, and M O'Loughlin,
\ajou Phys. Rev. &D45 (92) 3607.}
\lref\BaOl{T. Banks and M. O'Loughlin, \ajou Phys. Rev. &D47 (93) 540.}
\lref\DXBH{S.B. Giddings and A. Strominger,\ajou
Phys. Rev. &D46 (92) 627,.}
\lref\Hawktd{S.W. Hawking,\ajou Phys. Rev. Lett. &69 (92) 406.}
\lref\StTr{A. Strominger and S. Trivedi, ``Information consumption by
Reissner-Nordstrom black holes,'' ITP/Caltech preprint
NSF-ITP-93-15=CALT-68-1851, hep-th/9302080.}
\lref\SuTh{L. Susskind and L. Thorlacius,\ajou Nucl. Phys. &B382 (92) 123.}
\lref\dray{A. Ashtekar and T. Dray, \ajou Comm. Math. Phys. &79 (81) 581;
T. Dray, \ajou Gen. Rel. Grav. &14 (82) 109.}
\lref\dgt{H.F. Dowker, R. Gregory and J. Traschen, \ajou Phys. Rev. &D45 (92)
2762.}
\lref\HoWi{C.F.E. Holzhey and F. Wilczek, \ajou Nucl. Phys. &B380 (92) 447.}
\lref\unwald{W.G. Unruh and R.M. Wald,\ajou Phys. Rev. &D29 (84) 1047. }
\lref\rotbh{ A. Sen, \ajou Phys. Rev. Lett. &69, (92) 1006; J. Horne
and G. Horowitz \ajou  Phys. Rev.  &D46 (92) 1340.}
\lref\GGS{D. Garfinkle, S.B. Giddings and A. Strominger, ``Entropy in Black
Hole Pair Production'',
gr-qc/9306023, to appear in Phys. Rev. D.}
\lref\kw{W. Kinnersley and M. Walker, \ajou Phys. Rev. &D2 (70) 1359.}
\lref\senny{S. Hassan and A. Sen, \ajou Nucl. Phys. &B375 (92) 103.}
\lref\he{J. Ehlers, in {\it Les Theories de la Gravitation},
(CNRS, Paris, 1959); B. Harrison, \ajou J. Math. Phys. &9 (68) 1744.}
\lref\CGHS{C.G. Callan, S.B. Giddings, J.A. Harvey and A. Strominger,
\ajou Phys. Rev. &D45 (92) R1005.}
\lref\ebh{S.B. Giddings and A. Strominger, \ajou Phys. Rev. &D46 (92) 627.}
\lref\GH{G. W. Gibbons and J. B. Hartle, Phys. Rev. {\bf D42} (1990) 2458.}
\lref\kkbhs{H. Leutwyler, \ajou Arch. Sci. &13 (60) 549;
P. Dobiasch and D. Maison, \ajou Gen. Rel. and Grav. &14 (82) 231;
A. Chodos and S. Detweiler, \ajou Gen. Rel. and Grav. &14 (82) 879;
D. Pollard, \ajou J. Phys. A &16 (83) 565;
G.W. Gibbons and D.L. Wiltshire, \ajou Ann.Phys. &167
(86) 201;
erratum \ajou ibid. &176 (87) 393.}
\lref\gm{G.W. Gibbons and K. Maeda,
\ajou Nucl. Phys. &B298 (88) 741.}
\lref\ghs{D. Garfinkle, G. Horowitz, and A. Strominger,
\ajou Phys. Rev. &D43 (91) 3140, erratum\ajou Phys. Rev.
& D45 (92) 3888.}
\lref\AfMa{I.K. Affleck and N.S. Manton,
\ajou Nucl. Phys. &B194 (82) 38.}
\lref\alv{I.K. Affleck, O. Alvarez, and N.S. Manton,
\ajou Nucl. Phys. &B197 (82) 509.}
\lref\geroch{R.P. Geroch, \ajou J. Math. Phys. &8 (67) 782.}
\lref\raftop{R.D. Sorkin, in
{\sl Proceedings of the Third Canadian Conference on General
Relativity and Relativistic Astrophysics}, (Victoria, Can\-ada, May 1989),
eds. A. Coley, F. Cooperstock and B. Tupper (World Scientific, 1990).}
\lref\wheeler{J. Wheeler, \ajou Ann. Phys. &2 (57) 604.}
\lref\gwg{G.W. Gibbons,
in {\sl Fields and geometry}, proceedings of
22nd Karpacz Winter School of Theoretical Physics: Fields and
Geometry, Karpacz, Poland, Feb 17 - Mar 1, 1986, ed. A. Jadczyk (World
Scentific, 1986).}
\lref\CBHR{S.B. Giddings, ``Constraints on black hole remnants,'' UCSB preprint
UCSBTH-93-08, hep-th/9304027, to appear in {\sl Phys. Rev D}.}
\lref\garstrom{D. Garfinkle and A. Strominger,
\ajou Phys. Lett. &256B (91) 146.}
\lref\ernst{F. J. Ernst, \ajou J. Math. Phys. &17 (76) 515.}
\lref\rafKK{R. Sorkin, \ajou Phys. Rev. Lett. &51 (83) 87.}
\lref\grossperry{D. Gross and M.J. Perry, \ajou Nucl. Phys. &B226 (83) 29.}
\lref\hawk{S.W. Hawking in {\sl General relativity : an Einstein centenary
survey}, eds. S.W. Hawking, W. Israel (Cambridge University Press, Cambridge,
New York, 1979).}
\lref\schwinger{J. Schwinger,\ajou Phys. Rev. &82 (51) 664.}
\lref\melvin{M. A. Melvin, \ajou Phys. Lett. &8 (64) 65.}
\lref\Witt{E. Witten, \ajou Nucl. Phys. &B195 (82) 481.}
\lref\BOS{T. Banks, M. O'Loughlin and A. Strominger, \ajou Phys. Rev.
&D47 (93) 4476.}
\lref\ginsperry{P. Ginsparg and M.J. Perry, \ajou Nucl.Phys. &B222 (83) 245.}
\lref\Triv{S. Trivedi,\ajou Phys. Rev. &D47 (93) 4233.}
\lref\ChFu{S.M. Christensen and S. Fulling, \ajou Phys. Rev. &D15 (77) 2088.}
\lref\BGHS{B. Birnir, S.B. Giddings, J.A. Harvey and A. Strominger,
\ajou Phys. Rev. &D46 (92) 638.}
\lref\DGKT{H.F. Dowker, J.P. Gauntlett, D.A. Kastor and J. Traschen,
``Pair Creation of Dilaton Black Holes", hep-th/9309075,
to appear in Phys. Rev. {\bf D}.}
\lref\DGGH{H.F. Dowker, J.P. Gauntlett, G. T. Horowitz and S. B. Giddings,
``On Pair Creation of Extremal Black Holes and Kaluza-Klein Monopoles",
hep-th/9312172, submitted to Phys. Rev. {\bf D}.}
\lref\Ross{S. F. Ross, ``Pair Production of Black Holes in a $U(1)\otimes
U(1)$ Theory", hep-th/9401131.}
\lref\Cilar{S.B. Giddings, ``Comments on Information Loss and Remnants",
UCSBTH-93-35, hep-th/9310101.}
\lref\roternst{J.F. Plebanski and M. Demianski, \ajou Ann. Phys.
&98 (76) 98.}
\lref\bril{D. Brill, \ajou Phys. Rev. &D46 (92) 1560.}
\lref\kt{D.A. Kastor and J. Traschen,
``Particle Production and Positive Energy Theorems for Charged Black
Holes in De Sitter'',  UMHEP-399,  gr-qc/9311025, November 1992.}
%
%
\Title{\vbox{\baselineskip12pt
\hbox{EFI-94-17}
\hbox{hep-th/xxxxxxx}}}
{\vbox{\centerline{Pair Creation of Black Holes }
       }}
{
\baselineskip=12pt
\centerline{
Jerome P. Gauntlett}
\bigskip
\centerline{\sl Enrico Fermi Institute, University of Chicago}
\centerline{\sl 5640 S. Ellis Avenue, Chicago, IL 60637 }
\centerline{\it Internet: jerome@yukawa.uchicago.edu}
\medskip
\centerline{\bf Abstract}
This article is based on a talk given at the
IInd International Colloquium on Modern Quantum Field Theory, Bombay 1994.
The Ernst solution of dilaton gravity describes charged black holes
undergoing uniform accleration in a background magnetic field.
By analytically continuing the Ernst solution one obtains
instantons that
describe the pair production of black holes in the background field.
We review various aspects of these solutions paying special
attention to the Einstein-Maxwell, low-energy
string and $d=5$ Kaluza-Klein theories.
It is based on
work done in collaboration with
{}F. Dowker, S. Giddings, G. Horowitz, D. Kastor and J. Traschen
\refs{\DGKT,\DGGH}.
}



%
%

\newsec{Introduction}
It was first demonstrated by Schwinger \schwinger\ that
a uniform electric field is unstable to the formation of
electron-positron pairs.
This was generalised by Affleck and Manton \AfMa\ who showed that a
background magnetic field will spontaneously produce
monopole-antimonopole pairs.
In general relativity, the
analogue of a monopole is a magnetically charged black hole, and the
question
naturally arises as to whether black holes can be pair produced by a
background magnetic field.

Since the configuration of two black holes has a different
spatial topology than the vacuum, unlike the monopole case,
one cannot continuously deform one into
the other: pair production of black holes necessarily involves topology
change. The most natural framework for quantum
gravity in which to investigate such processes is consequently
the sum-over-histories formulation. We assume that the path-integrals
can be evaluated semi-classically using instanton techniques and thus
we look for
finite action solutions to the euclidean equations of motion that
interpolate between the appropriate initial and final 3-geometries.

Affleck and Manton used an approximate instanton to estimate the pair
creation rate for monopoles (see also \alv).
As Gibbons first realized \gwg,
an exact instanton for the Einstein-Maxwell theory can be obtained
by analytically continuing a solution
found by Ernst almost twenty years ago \ernst. The Ernst solution describes two
oppositely charged black holes undergoing uniform acceleration in a background
magnetic
field. (Ernst actually considered electric fields, but by duality,
that is equivalent to the magnetic fields we will consider here.)
The Ernst solution describes the evolution of the black holes after their
creation. Regularity of the euclidean instanton turns out to
restrict the charge to mass
ratio of the black holes. Gibbons believed that only extremal black holes
could be created. But
Garfinkle and Strominger \garstrom\ found a regular instanton for which the
black holes were slightly non-extremal and
the horizons of the two black holes were identified
to form a wormhole in space. The action of the instanton was evaluated
in \garstrom\ to give the rate of production of the wormhole instantons
and it was shown that the leading term was simply the Schwinger rate.

In more recent work these investigations have been extended in the context
of ``dilaton gravity" describing the interaction between a dilaton,
a $U(1)$ gauge field and gravity via the action
\eqn\action{
S={1\over 16 \pi}\int d^4x
{\sqrt {-g}}\left[R-2(\nabla\phi)^2-e^{-2a\phi}F^2\right].
}
{}For $a=0$ it is always consistent with the equations of
motion to set $\phi=0$ and we recover
the standard Einstein-Maxwell theory.
{}For $a=1$, $S$ is part of the action
describing the low energy dynamics of string theory.
Note that for some physical
questions it is more appropriate to use the conformally rescaled
string metric,
$\tilde g_{\mu\nu} = e^{2\phi} g_{\mu\nu}$.
The value $a=\sqrt{3}$ is also of special interest since this corresponds to
standard
Kaluza-Klein theory, i.e. \action\ is equivalent to the
five-dimensional vacuum Einstein action for geometries with a spacelike
symmetry if we define the five dimensional metric via
\eqn\fivmet{ ds^2 = e^{-4\phi/\sqrt 3}( dx_5 + 2A_\mu dx^\mu)^2
     +e^{2\phi/\sqrt 3} g_{\mu\nu} dx^\mu dx^\nu \ . }

{}For all values of $a$ there exist static charged black hole solutions which
we shall briefly review below.
In \DGKT\ the generalisation of the Ernst
solution was constructed for all values of $a$ and it was shown that
regular wormhole type instantons exist only for $a<1$.
However, it was shown in \DGGH\ that extremal instantons exist for all values
of
$a$. Furthermore the action for both type of
instantons was calculated in \DGGH\ and again it was shown that
the rate of pair production, to leading order, is the Schwinger rate.
This article is essentially a summary of the results of \DGKT\ and \DGGH.

The plan of the rest of the article is as follows. In section 2 we review
the black hole solution and discuss
the analogue of a uniform magnetic field in General Relativity,
the Melvin solution.
The dilaton Ernst solution
is discussed in section 3 and in particular we examine the extremal limit.
In section 4 we discuss the euclidean instantons obtained
by analytic continuation of the Ernst solution
and calculate their action.
In section 5 we make some coments on quantum corrections
and section 6 contains some conclusions.

\newsec{Background}
\subsec{Black Hole Solutions}
The static, spherically symmetric magnetically
charged black hole solution is given by
\refs{\gm,\ghs}
\eqn\dbhs{
\eqalign{
& ds^2=-\lambda^2dt^2+\lambda^{-2}dr^2+R^2(d\theta^2 +
\sin^2\theta d\varphi^2)
 \cr
 & e^{-2a\phi}=\left(1-{r_-\over r}\right)^{2a^2\over(1+a^2)},\qquad
 \qquad A_\varphi=q(1 - {\rm cos}\theta)\cr
 &\lambda^2=\left(1-{r_+\over r}\right)
 \left(1-{r_-\over r}\right)^{{(1-a^2)\over (1+a^2)}},\qquad
 R^2=r^2\left(1-{r_-\over r}\right)^{2a^2\over (1+a^2)}.\cr}
 }
If $r_+ > r_-$, the surface $r=r_+$ is the event horizon.
{}For $a=0$, the surface $r=r_-$ is the
inner Cauchy horizon of the Reissner-Nordstrom black hole, however for $a>0$
this  surface is singular.
The parameters $r_+$ and $r_-$ are related to the ADM mass $m$
and total charge $q$ by
\eqn\mass{
m={r_+\over 2} + \left ({1-a^2\over 1+a^2}\right ){r_-\over 2},\qquad
q=\left({r_+r_-\over 1+a^2}\right)^{1\over 2}.}

The extremal limit occurs when $r_+=r_-$.
The spatial geometry of the extreme Reissner-Nordstrom black hole
resembles an infinite throat connected onto an asymptotically flat region.
The situation in low-energy string theory is similar: the extremal
spatial geometry is the same as long as one uses the string metric.
In addition the entire extremal string 4-geometry is non-singular.
In the Kaluza-Klein case the extremal geometry is just the Kaluza-Klein
monopole \refs{\rafKK, \grossperry} : the five dimensional metric \fivmet\
is the product
of euclidean Taub-Nut with a real line and is non-singular.

\subsec{Dilaton Melvin Spacetimes}
The concept of a uniform
magnetic field in flat spacetime must be modified when gravity is
included since the field has non-zero stress-energy. The natural
generalisation in Einstein-Maxwell theory is the Melvin solution \melvin\
which represents an infintely long straight flux-tube. The solution
is axisymmetric and static, the gravitational
attraction being balanced
by the transverse magnetic pressure.

The generalisation of the Melvin solution to dilaton gravity is given by
\gm
\eqn\dmelv{
\eqalign{
&ds^2=\Lambda^{2\over 1+a^2}\left[-dt^2+d\rho^2+dz^2\right]
+\Lambda^{-{2\over 1+a^2}}\rho^2d\varphi^2\cr
&e^{-2a\phi}=\Lambda^{2a^2\over 1+a^2},\qquad
A_\varphi=-{2\over (1+a^2)B\Lambda}\cr
&\qquad \Lambda=1+{(1+a^2)\over 4}B^2\rho^2\cr}
}
and
now both the gravitational and scalar attraction balance the magnetic pressure.
The square of the Maxwell field is $F^2 = 2B^2/\Lambda^4$,
which is a maximum
on the axis $\rho=0$ and decreases  to zero at infinity. The parameter
$B$ labels the strength of the magnetic field.

\newsec{Dilaton Ernst}
\subsec{General Properties}
The generalisation of the Ernst metric of Einstein-Maxwell theory
to dilaton gravity was constructed in \DGKT\
and is given by
\eqn\dernst{
\eqalign{
&ds^2=(x-y)^{-2}A^{-2}\Lambda^{2\over 1+a^2}
\left[F(x)\left\{G(y)dt^2-G^{-1}(y)dy^2\right\}
+F(y)G^{-1}(x)dx^2\right]\cr &\qquad +
(x-y)^{-2}A^{-2}\Lambda^{-{2\over 1+a^2}}F(y)G(x) d\varphi^2\cr
&e^{-2a\phi}=e^{-2a\phi_0}\Lambda^{2a^2\over 1+a^2}
{F(y)\over F(x)},\quad
A_\varphi=-{2e^{a\phi_0}\over (1+a^2)B\Lambda}\[1+{(1+a^2)\over 2}Bqx\]
\cr
}
}
where the functions $\Lambda \equiv\Lambda(x,y)$,
 $F(\xi)$ and $G(\xi)$ are given by
\eqn\fns{\eqalign{
&\Lambda=\[1+{(1+a^2)\over 2}Bqx\]^2+{(1+a^2)B^2\over 4A^2(x-y)^2}
G(x)F(x) \cr
&F(\xi)=(1+r_-A\xi)^{2a^2\over (1+a^2)}\ \cr
&G(\xi)=(1-\xi^2-r_+A\xi^3)(1+r_-A\xi)^{(1-a^2)\over
(1+a^2)}\ .\cr
}}
and $q$ is related to $r_+$ and $r_-$ by \mass. As we will discuss,
this solution possesses both an inner and outer black hole
horizon in addition to an acceleration horizon and asymptotically
approaches the Melvin solution. Consequently, we can interpret the solution
as describing charged black holes being uniformly accelerated in a
background magnetic field.

The constant $\phi_0$ in the solution for the
dilaton determines the value of the dilaton at infinity. Although one could
keep this as a free parameter, we will fix it so that the dilaton vanishes
on the axis at infinity in agreement with \dmelv.
The solution \dernst\ depends
on four other parameters, $r_\pm,A,B$. Defining $m$ and $q$ via \mass\
we can loosely
think of these parameters together with $A,B$
as denoting the mass, charge and acceleration
of the black
holes and the strength of the magnetic field which is accelerating them,
respectively.
We emphasize, however, that this is heuristic since, for
example, the mass and acceleration are not in general
precisely defined and,
further, we will see that $q$ and $B$ only approximate the physical
charge $\widehat q$ and magnetic field $\widehat B$ in the limit
$r_\pm A\ll1$.

Before discussing the causal structure of the solution,
it is convenient to introduce the following notation.
Define $\xi_1\equiv -{1\over r_-A}$ and let $\xi_2\le\xi_3<\xi_4$
be the three roots  of the cubic in $G$.
We restrict the range of the parameters $r_+$ and $A$ so that
$r_+A\le2/(3\sqrt{3})$,
so that the $\xi_i$ are all real; the limit $r_+A=2/(3\sqrt{3})$
corresponds to $\xi_2=\xi_3$.
We also restrict the parameter $r_-$
so that $\xi_1 \le \xi_2$.

The metric \dernst\ has two Killing vectors,
${\partial\over\partial t}$ and ${\partial\over\partial\varphi }$.
The surface $y=\xi_1$ is
singular for $a>0$, as can be seen from the square of the field strength.
This surface is analogous to the singular surface
(the ``would be'' inner horizon) of the dilaton black holes.
The surface $y=\xi_2$ is the black hole horizon and the surface
$y=\xi_3$ is the acceleration horizon; they are both Killing horizons for
${\partial\over\partial t}$.

The coordinates $(x,\varphi)$ in \dernst\ are angular coordinates.
To keep the signature of the metric fixed, the coordinate $x$
is restricted to the range $\xi_3 \le x\le\xi_4$ in which $G(x)$ is positive.
Due to the conformal factor $(x-y)^{-2}$ in the metric,
spatial infinity is reached by fixing $t$ and
letting both $y$ and $x$ approach
$\xi_3$.  Letting $y\rightarrow x$ for $x\ne\xi_3$ gives null or
timelike infinity \dray .
Since $y\rightarrow x$ is infinity, the range of the coordinate $y$ is
$-\infty<y<x$ for $a=0$, $\xi_1<y<x$ for $a>0$.

The norm of the
Killing vector ${\partial\over\partial\varphi }$ vanishes at $x=\xi_3$ and
$x=\xi_4$, which correspond to the poles of the spheres surrounding the black
holes.   The axis $x=\xi_3$ points along the symmetry axis
towards spatial infinity.  The axis $x=\xi_4$ points towards the other black
hole. Note that the coordinates we are using only cover one region of
spacetime containing one of the black holes.
As discussed in \DGKT,
to ensure that the metric is free of conical singularities at both poles,
$x=\xi_3, \xi_4$, for a single choice of period for $\vp$,
we must impose the condition
\eqn\nonodes{G^\prime(\xi_3)\Lambda(\xi_4)^{2\over 1+a^2}
= -
G^\prime(\xi_4)\Lambda(\xi_3)^{2\over 1+a^2}.}
where\foot{It follows from \fns\ that
when $x$ is equal to
a root of $G(x)$, $\Lambda(x,y)$ is independent of $y$.  So $\Lambda(\xi_i)$
are constants.} $\Lambda(\xi_i)\equiv \Lambda(x=\xi_i)$.
When \nonodes\ is satisfied, the spheres are regular as long as $\vp$
has period
\eqn\phiperiod{\Delta\vp={4\pi \Lambda^{2\over 1+a^2}(\xi_3)\over
G^\prime(\xi_3) }\ .}
The condition \nonodes\
can be readily understood in the limit $r_\pm A\ll 1$
where it becomes
Newton's law,
\eqn\Newt{mA\approx qB\ ,}
where we have used \mass\ to replace $r_\pm$ with $m,q$. This is true for
all $a$. More generally,
the condition \nonodes\ reduces the number of free
parameters for the solution to three  by relating the
acceleration to the magnetic field, mass, and charge.

To show that the Ernst solution approaches the Melvin solution
at large
spacelike distances,
$x,y \rightarrow \xi_3$,
we change coordinates from $(x,y,t,\vp)$ to $(\rho,\zeta,\eta,\phiT)$ using
\eqn\coordch{\eqalign{&x-\xi_3 = {4F(\xi_3) \Lambda^{2\over 1+a^2}(\xi_3)
\over G'(\xi_3) A^2} {\rho^2\over (\rho^2+\zeta^2)^2},\quad
\xi_3-y
={4F(\xi_3) \Lambda^{2\over 1+a^2}(\xi_3)
\over G'(\xi_3) A^2} {\zeta^2\over (\rho^2+\zeta^2)^2} \cr
&t= {2\eta\over  G'(\xi_3)}\quad ,\quad \vp = {2 \Lambda^{2\over 1+a^2}(\xi_3)
\over G'(\xi_3) }\phiT\ }}
Note that  $\eta, \phiT$  are related to $t,\vp$ by a simple rescaling and
that $\phiT$ has period $2\pi$ due to \phiperiod.
Choosing the constant $\phi_0$ appropriately,
for large $\rho^2+\zeta^2$, the dilaton Ernst metric reduces to
\eqn\blip{
ds^2  \to \tilde\Lambda^{2\over 1+a^2}\left(-\zeta^2d
\eta^2+d\zeta^2+d\rho^2\right)+\tilde\Lambda^
{-{2\over 1+a^2}}\rho^2 d\widetilde\varphi^2}
where
\eqn\blob{
\eqalign{&\tilde\Lambda=\left(1+{1+a^2\over 4}\widehat B^2\rho^2\right)\
,\cr
&\widehat B^2=
{B^2G^{\prime}(\xi_3)^2\over 4\Lambda^{3+a^2\over 1+a^2}(\xi_3)}\ .}
}
which we identify as the dilaton Melvin metric in Rindler-type coordinates.
In particular we note the coordinate $t$ in the dilaton Ernst solution
is the analogue of Rindler time (and hence, since ${\p\over \p t}$ is a Killing
vector the spacetime is boost invariant, not static).
It is clear that the physical magnetic charge is $\widehat B$.
The physical charge of the black hole is defined by $\hq = {1\over 4 \pi}
\int F$ where the integral is over any two sphere surrounding the black hole.
In the weak field limit $r_\pm A\ll 1$, $\widehat B\approx B$ and $\hq \approx
q$.

\subsec{The Limit $\xi_1=\xi_2$: Accelerating Extremal Black Holes}

Since $y=\xi_2$ is the event horizon and $y=\xi_1$ is an inner horizon
($a=0$) or singularity ($a >0$), it follows that the extremal limit
of the dilaton Ernst solution is given by choosing the parameter $r_-$
 so that $\xi_1=\xi_2$. Recalling the regularity condition \nonodes,
it follows that the extremal solution is described by two
parameters which we can take to be the physical charge $\hq$ and
magnetic field $\widehat B$.
It was shown in detail in \DGGH\ that
that as one approaches the horizon $y\rightarrow \xi_2$
the extremal solution becomes spherically symmetric, and approaches the
static black hole  solution \dbhs\ with $r_-=r_+$. This surprising
result has
a number of consequences which we now discuss.

{}First, all
the geometric properties of the extremal static solution near the horizon
carry over immediately to the accelerated case. In particular,
for $a=0$, a constant-$t$ slice of the solution has
an infinitely long throat. {}For $a=1$, the string metric
$d\tilde s^2 = e^{2\phi} ds^2$ also has an infinite throat in which
the solution takes the form of the linear dilaton
vacuum. {}For $a=\sqrt{3}$,
the five dimensional metric \fivmet\ approaches that of the Kaluza-Klein
monopole.

A second consequence is that there is a sense in which the extremal
black holes are not accelerating. {}For $a=0$, this is suggested by the
fact that the event horizon is exactly spherical. But a more convincing
argument comes from examining the acceleration of a family of observers
near the horizon whose four velocities are proportional to $\p /\p t$.
{}For the static black hole, the acceleration of these observers approaches
the finite limit $1/q$ as they approach the horizon. (This is
related to the fact that the surface gravity vanishes for extremal
black holes and is in contrast to the non-extremal case in
which the acceleration diverges.)
If one computes the acceleration of these observers for the Ernst solution,
one again finds that it approaches $1/\widehat q$ as $y \rightarrow \xi_2$
independent of direction. Although this particular argument cannot be
extended to $a >0$ since the acceleration (in the Einstein metric)
now diverges for the static
extremal solution
near the horizon, other arguments can be made. {}For example,
when $a=1$, in the string metric the acceleration of these observers
tends to zero down the throat. In addition,
when $a=\sqrt{3}$, $y=\xi_2$ is a regular
origin in the five dimensional Kaluza-Klein solution, and one can show that
its worldline is a geodesic!

Even though the black hole itself is not accelerating, the region around
the black hole is. This is clear from the relation between the  solution
and the dilaton Melvin solution in accelerating coordinates discussed in
the previous subsection. In terms of the infinite throats, one might say
that the mouth of the throat is accelerating  while the region down the
throat is not.

\newsec{Dilaton Ernst Instantons}
To calculate tunnelling effects using instanton methods
one looks for a classical solution to the euclidean equations of motion that
interpolates between the appropriate initial and final 3-geometries.
In addition, the classical
solution must have zero momentum (zero extrinsic curvature)
at the initial and final configurations.
Let us call such a solution a ``tunneling geometry" \GH.
By analytically continuing the
Ernst solution we obtain instantons or more precisely ``bounces"
that have a moment of time symmetry.
The corresponding tunneling geometries are obtained by slicing the
manifold along the moment of time symmetry. By construction it is clear
that the final geometry has vanishing extrinsic curvature and
furthermore that
the final geometry provides good initial data for the subsequent lorentzian
evolution given by the Ernst solution. To leading
order in the semi-classical expansion, the tunnelling rate is given by
$e^{-S_E}$ where $S_E$ is the euclidean action of the instanton after
subtracting off the action of the background, in our case the background
magnetic field.

Euclideanizing \dernst\ by setting $\tau=it$, we find that
another condition must be imposed
on the parameters
in order to obtain a regular solution. Two distinct ways that
this may be achieved were discussed in \DGKT\ and are reviewed
in the first subsection below. These include the wormhole instantons.
There is a third
way \DGGH\ which leads to  the extremal instantons
and is described in the second subsection. The calculation of the
action for the wormhole and extremal instantons is given in the
third subsection.

\subsec{Wormhole Instantons}

In the lorentzian solution, the vector $\p/\p t$ is timelike only for
$\xi_2 <y < \xi_3$.
If we make the restriction
$\xi_1<\xi_2$
then the Einstein metric has a regular horizon for all values of $a$.
In this case,
one must impose a condition on the parameters in
order to
eliminate conical singularities in the euclidean solution
at both the black hole ($y=\xi_2$) and
acceleration ($y=\xi_3$) horizons
with a single choice of the period of $\tau$.
This is equivalent to demanding that the
Hawking temperature of the black hole horizon equal the Unruh
temperature of the acceleration horizon.

In terms of the metric function $G(y)$ appearing in \dernst, the period of
$\tau$ is taken to be
\eqn\tauper{\Delta \tau ={4\pi\over G^\prime(\xi_3)}}
and the constraint
is
\eqn\regular{
G^\prime(\xi_2) =- G^\prime(\xi_3), }
yielding
\eqn\roots{
\left({\xi_2-\xi_1\over \xi_3-\xi_1}\right)^{1-a^2\over 1+a^2}(\xi_4-\xi_2)
(\xi_3 -\xi_2)
=(\xi_4 - \xi_3)(\xi_3 -\xi_2). }
With $\xi_1<\xi_2$ there are two ways to satisfy this condition
and correspondingly two types of instantons.
The first one exists when $\xi_2\ne\xi_3$ and only
for $0\le a<1$. It has topology $S^2 \times S^2 - \{pt\}$
where the removed point is $x=y=\xi_3$.
This instanton is readily interpreted as a bounce: the surface defined by
$\tau=0,\Delta\tau/2$ has topology
$S^2 \times S^1 - \{pt\}$, which is that
of a wormhole attached to a spatial slice of Melvin and is the zero momentum
initial data for the lorentzian Ernst solution.
The tunneling geometry
describes the pair creation of a pair of oppositely charged dilaton
black holes in a magnetic field which subsequently uniformly accelerate
away from each other. {}From the metric
we deduce that there is a horizon sitting inside
the wormhole throat, located at a finite proper distance from the mouth.

These ``wormhole'' instantons generalize
the Einstein-Maxwell instanton
discussed in  \garstrom. The reason these instantons
only exist for $0\le a<1$ can be understood (for weak fields) by recalling the
thermodynamic behavior
of the dilaton black holes as extremality is
approached: the Hawking temperature, as defined from the
period of $\tau$ in the euclidean section, goes to zero for $0\le a<1$,
approaches a constant for $a=1$ and diverges for $a>1$.
Thus, for small magnetic fields and hence accelerations,
we expect to be able to
match the resultant Unruh temperature and the black hole temperature
by a small perturbation of the black
hole away from extremality only for $0\le a<1$.

The second class of instantons
we mention only for completeness since their interpretation
is obscure. They are defined by
$\xi_2 = \xi_3$ which is equivalent to $r_+A = 2/(3\sqrt{3})$,
and
have topology $S^2 \times R^2$.
Note that for these instantons one does not
have to impose the condition \nonodes\ for regularity.

\subsec{Extremal Instantons}

The wormhole type instantons discussed above were made regular
by the condition that the temperatures of the black hole
and acceleration horizons should be equal.
Gibbons \refs{\gwg} pointed out (for $a=0$) that
there
is another way that the temperatures of the black hole
and acceleration horizons can be equal: that is if the
black hole is extremal. This might seem strange since
the extremal Reissner-Nordstrom black hole has zero temperature in the
sense that the euclidean time coordinate need not be periodically
identified to obtain a regular geometry. But, of course we
{\it{can}} periodically identify the euclidean time and with any period we
like (just as for flat space). In particular we choose $\tau$ to have period
\tauper\ to
eliminate the conical singularity at the acceleration horizon, $y=\xi_3$.

{}For $a=0$ the extremal condition $\xi_1=\xi_2$ does indeed lead
to a smooth instanton. The range of the
coordinate $y$
is $\xi_2< y\le \xi_3$ in the euclidean section.
We noted in Section 3.2 that the lorentzian
solution near the back hole is just that of an extremal
black hole. The same holds for the euclidean solution.
The horizon $y=\xi_2$ is infinitely far away (in every
direction since every direction is now spacelike) and gives
no restriction on the period of $\tau$. Thus we have
obtained a regular geometry with internal infinities down the
throats of the extremal black holes. The length of the $y=$ constant
circles in the $(y,\tau)$ section tend to zero
as $y \to \xi_2$
but the curvature remains bounded and the radii of the two spheres
approach a constant.
The topology of this instanton is
$R^2 \times S^2 - \{pt\}$ with the removed point again being $x=y=\xi_3$.
The 3-geometry created by the tunneling geometry,
the $\tau=0,\Delta\tau/2$
zero momentum slice, is a spatial slice of a Melvin universe
with two infinite tubes attached.

The extremal case $\xi_1=\xi_2$ also gives well defined
instantons for $0<a\le 1$. Although the Einstein
metric has a  singularity, the so called ``total" metric \gm,
$ds_T = e^{2\phi\over a} ds^2$, which is the same as the
string metric for $a=1$, is perfectly regular.
We noted in Section 3.2 that the
metric close to the singularity is that of the extremal
black hole. In the total metric this looks like
\eqn\lindil{
ds_T^2 \propto -d{t}^2 + \left(1-{r_+\over r} \right)^{-{4\over 1+a^2}}dr^2
+{r_+}^2 \left(1-{r_+\over r} \right)^
{{2(a^2-1)\over 1+a^2}}
d\Omega_2^2
}
{}For $0<a<1$,
the total metric is geodesically complete and the
spatial sections have the form of two asymptotic regions joined by a
wormhole, one region being flat, the other having a deficit solid
angle. Hence the corresponding extremal instantons are regular
as long as we again choose the period of $\tau$ to be \tauper\ to
ensure regularity as $y\to \xi_3$.
{}For $a=1$ the geometry of the string metric
is that of an infinitely long throat
of constant radius
and thus
the $a=1$ extremal instanton
looks very much like that of the $a=0$
extremal instanton described above: the topology is the
same, $R^2 \times S^2 - \{pt\}$, and the major difference
is that the proper radius of $y=$constant circles in the $(y,\tau)$
section tends to
a finite limit as $y\rightarrow \xi_2$. The $\tau=0,{\Delta\tau\over2}$
slice resembles the $a=0$ case and hence the tunneling geometry describes the
pair production of extreme $a=1$ black holes with their infinite throats
(in the string metric).
It is perhaps worth pointing out that as $a\to 1$, the wormhole instanton
approaches the $a=1$ extremal instanton.

{}For $a>1$, both the Einstein metric and the
total metric have a  naked singularity in the extremal limit.
It has been argued in \HoWi, however, that these
``black holes" should be interpreted
as elementary particles. The extremal instantons can then be
interpreted as pair creating such objects.
{}For $a=\sqrt{3}$, the five dimensional metric of the extremal instanton
is already regular at $y=\xi_2$ and hence
is regular if $\tau$ has period \tauper.
It was shown in \DGGH\ that the topology of the instanton is
$S^5-S^1$ and that the topology of the zero momentum slice is $S^4-S^1$.
It describes the creation of a Kaluza-Klein monopole-anti-monopole pair.

\subsec{The action}

To leading semiclassical order, the pair production rate of non-extremal or
extremal black holes is given by
$e^{-S_E}$ where $S_E$ is the euclidean action of the corresponding
instanton solutions.
The euclidean action including boundary terms is given by
\eqn\eac{
S_E={1\over 16\pi }\int_V d^4x{\sqrt g}\left[-R+2(\nabla\phi)^2+
e^{-2a\phi}F^2\right]-{1\over 8\pi }\int_{\partial V} d^3x{\sqrt h}K
}
where $h$ is the induced three metric
and $K$ is the trace of the extrinsic curvature of the boundary.

{}For both the wormhole and extremal instantons there is a boundary
at infinity,
$x=y=\xi_3$ which contributes an infinite amount to the action.
However, the action of the background magnetic field solution is itself
infinite. It was shown in \DGGH\ how to subtract
the infinite  background contribution to
obtain the physical result.
{}For the extremal instantons there is also an additional
boundary down the throats of the black holes i.e. at $y=\xi_2$.
The contribution to the action from this boundary vanishes.

The result of the calculations of \DGGH\ is that
the action is finite for both types of
instantons and is given by
\eqn\cta{
S_E=
2\pi \hq^2 {\Lambda(\xi_4)(\xi_3-\xi_2)\over \Lambda(\xi_3)(\xi_4-\xi_3)}.
}
{}For the $a=0$ wormhole instantons this agrees
with the result first obtained (using a different method) by
Garfinkle and Strominger \garstrom.
Notice that the result is finite for the extremal instantons
despite the infinite throats for $0\le a\le 1$ and despite the
fact that there are
singularities in both the Einstein and the total metric
for $a>1$.
The action can be expressed in terms of the physical charge
$\widehat q$ and magnetic field $\widehat B$ by
expanding out in the parameter $\widehat q\hB$.
The action for the wormhole type instantons is
\eqn\wormacta{
\eqalign{
&S_E=\pi\widehat q^2\[{1\over \widehat q\hB}-{1\over 2}
+\cdots\]
\qquad\qquad a=0\cr
&S_E=\pi\widehat q^2\[{1\over (1+a^2)\widehat q\hB}+{1\over 2}
 +\cdots \]
\qquad 0< a<1\cr}
}
while the action for the extremal type instantons for all $a$ is given by
\eqn\extact{
S_E=\pi\widehat q^2\[{1\over (1+a^2)\widehat q\hB}+{1\over 2}
+\cdots\]
}
where dots denote higher order terms which may be fractional
powers of  $\hq\hB$.
To leading order these all give the Schwinger result, $\pi m^2/
\widehat q\hB$ after using the relation between the
mass and charge of extremal black holes,
$(1+a^2)m^2=\widehat q^2$.

To next-to-leading order, for $a=0$ the action of
the extremal instanton is greater
than the action of the wormhole instanton
by $\pi {\widehat q}^2 = {1\over 4} A$  where $A$ is the area
of the horizon of an extremal black hole of charge $\widehat q$.
In fact, to this order it could also be the area of the horizon
of the wormhole instanton. This difference
is precisely the Bekenstein-Hawking
entropy.  {}For $0<a<1$ the difference is zero to this order, which
is consistent with the difference being the area of the
horizon of the extremal instanton since that vanishes for $a>0$. The
area of the horizon in the wormhole instanton is non-zero, but  higher order
in $\hq \hB$.

In \GGS\ a comparison was made between the wormhole action
 for $a=0$ and the action of an instanton describing the
creation of a monopole-anti-monopole pair.
It was found that the action of the monopole instanton
was greater than that of the wormhole instanton by
the black hole entropy. Our result thus suggests that,
at least for $a=0$, the extremal black holes behave more like
elementary particles than non-extremal ones.   However, these conclusions
neglect quantum corrections, to which we
now turn.

\newsec{Quantum Considerations}
Until now we have been discussing the solutions purely at the
classical level.  In this section we will make some general comments
about quantum corrections referring the reader to \DGGH\ for
more details.

Lets first consider the lorentzian solution with general $m$ and $q$.
An observer
travelling on a trajectory at a fixed distance
from the black hole will be accelerated\foot{This is of course in addition
to the usual acceleration needed to avoid falling into the black hole
were it static.}
and therefore would observe
acceleration radiation if carrying a detector.
This suggests that we should
describe the black hole as being in contact with this approximately thermal
radiation.  If so, then the black hole would be expected to absorb energy,
thus perturbing the solution. However, the black hole can
also emit Hawking radiation, and therefore achieve a time-independent
equilibrium state where the emission and absorption
rates match.
We have seen evidence that this might be possible
(at least for $0\le a<1$) in the construction of the euclidean wormhole
instantons: it is possible for $0\le a <1$ to constrain the parameters
via \regular\ to match
the Hawking and Unruh temperatures.

To address the issue of quantum corrections to the extremal geometries
it is crucial to know the forms of the effective potentials for
fluctuations about the extremal black holes. This was studied in detail
in
\refs{\HoWi,\DXBH}
and was shown to depend critically on the value of $a$.
{}For $a>1$
the potentials grow arbitrarily large as the horizon (singularity) is
approached suggesting that the quantum corrections do not move
the solution away from extremality.
However, for $0\le a\le 1$ there are potential
barriers outside the black hole, but these
vanish at the horizon (for the string theory case this requires the
addition of a particular
kind of matter to the action \action).
This suggests that the acceleration radiation
can perturb the solution away from extremality.
As noted above, there is a potential equilibrium solution
for $0\le a<1$ into which the extreme solution could evolve. However,
it is not at all clear what happens to the $a=1$ extremal geometry.
It is perhaps worth pointing out that the string coupling constant
$e^{2\phi}$ is getting large down the infinite throat and we expect
string loop effects to substantially modify the solution in this region.

Similar comments apply to the instanton solutions. Since the
Hawking and Unruh temperatures are matched for the wormhole
instantons we do not expect significant quantum corrections. {}For the extremal
instantons with $a>1$ we don't expect quantum corrections because of the
infinite potential barriers around the singularity. On the other hand we
expect significant quantum corrections to the extremal instantons
with $0\le a\le 1$. Are the extremal instanton solutions
with $0\le a<1$ somehow
perturbed into the wormhole solutions? This is not clear since the two types
of instantons pair produce black holes in a different topological class.
It is even less clear what happens to the extremal instantons in
string theory.

\newsec{Conclusions}
We have reviewed the Ernst solution of dilaton gravity which
describes two dilaton black holes being uniformly accelerated in
a uniform magnetic field. In the extremal limit,
as one approaches the black hole horizon the solution reduces exactly
to the static dilaton black hole solution. In this limit there is thus a
sense in which although the surrounding spacetime is accelerating
the black holes themselves are not.

By analytically continuing the Ernst solution one obtains two types
of finite action instantons.
The extremal instantons exist for all values of
$a$ and describe the pair creation of
extremal black holes. {}For $a=\sqrt 3$
the extremal instanton describes the pair creation of Kaluza-Klein monopoles.
The wormhole instantons exist for $0\le a<1$ and describe the
pair creation of two non-extremal black holes connected at their horizons
to form a wormhole.
Although we do not expect quantum corrections to significantly change the
geometry of the wormhole instantons or the extremal instantons with $a>1$,
we do expect significant corrections to the extremal instantons with $a\le 1$.

There is some evidence that $a=1$ wormhole
instantons exist if one allows the black holes to rotate.  One
piece of evidence comes from \BOS, where  an approximate
wormhole instanton was constructed
which includes rotation. Another comes from the
fact that for the rotating black hole with $a=1$, the
Hawking temperature goes to zero in the extremal limit whenever the angular
momentum is non-zero \rotbh. Thus one could match the Unruh temperature at the
acceleration horizon by a slightly non-extremal rotating black hole.
Note that a wormhole solution to the $a=1$ theory with 2 $U(1)$'s has recently
been constructed in \Ross.

One of the most important issues is to develop a better understanding
of quantum corrections to the instanton
approximation and their
effects on the geometry and pair creation rate.
It is particularly important to understand the
calulation of the rate,
as a finite answer may indicate that such
black holes serve as a model for black hole
remnants \refs{\CGHS,\BDDO,\DXBH}.
A better understanding
of these corrections will also help to resolve the question of
whether an infinite volume of space can really be created in a finite
amount of time. If so, there would appear to be problems with causality,
unless the state down the throats were fixed uniquely.
Hopefully I will have the opppurtunity of reporting on these
and other topics at a future colloquium.

\bigskip\centerline{\bf Acknowledgements}\nobreak
I would like to thank the conference organisers for a stimulating
and enjoyable conference.
This work was supported by a grant from the
Mathematical Discipline Center of the Department of Mathematics,
University of Chicago.

\listrefs

\end